\documentclass[useAMS,usenatbib]{mn2e}

\usepackage{graphicx}
\usepackage{rotating}
\usepackage{pdflscape}
\usepackage{amsmath}
\usepackage{url}

\title[{A multi-wavelength study of SXP 
1062}]{\centering A multi-wavelength study of SXP 
1062, the long period X-ray pulsar  associated with a 
supernova remnant\thanks{The 
scientific results reported in this article are based on observations made by 
the \chandra, \xmm, and \swift\ X-ray observatories, the Southern African Large 
Telescope (\salt), and the OGLE\,III photometry data base.}}
\author[A.~Gonz\'alez-Gal\'an et al.]
{A.~Gonz\'{a}lez-Gal\'{a}n$^{1}$, L.~M.~Oskinova$^{1,2}$\thanks{E-mail: 
lida@astro.physik.uni-potsdam.de}, S.~B.~Popov$^{3}$, 
 F.~Haberl$^{4}$,
M.~K\"{u}hnel$^{5}$, \newauthor J.~Gallagher III$^{6}$, 
M.~P.~E.~Schurch$^{7}$, 
M.~A.~Guerrero$^{8}$
\\
$^{1}$Institute for Physics and Astronomy, University of Potsdam, D-14476 Potsdam, Germany\\
$^{2}$Kazan Federal University, Kremlevskaya Str., 18, Kazan, Russia\\
$^{3}$Sternberg Astronomical Institute, Lomonosov Moscow State University, 
119992 Moscow, Russia\\
$^{4}$Max-Planck-Institut f\"{u}r extraterrestrische Physik, 
Giessenbachstra{\ss}e, 85748 Garching, Germany\\
$^{5}$Dr. Karl Remeis-Sternwarte \& ECAP, Universit\"at Erlangen-N\"urnberg, 
Sternwartstr. 7, 96049 Bamberg, Germany\\
$^{6}$ University of Wisconsin-Madison, 475 N. Charter St. Madison, WI 53706 
USA\\
$^{7}$Astrophysics, Cosmology and Gravity Centre (ACGC), Astronomy Department, 
University of Cape Town, Rondebosch,\\
Private Bag X1 7701, South Africa\\
$^{8}$Instituto de Astrof\'{i}sica de Andaluc\'{i}a, IAA-CSIC, Glorieta de la 
Astronom\'{i}a s/n, 18008 Granada, Spain\\
}

\def\gtsima{$\; \buildrel > \over \sim \;$}    
\def\gtrsim{\lower.5ex\hbox{\gtsima}}           
\def\lesssim{\lower.5ex\hbox{\ltsima}}           
\def\ltsima{$\; \buildrel < \over \sim \;$}    

\newcommand{\nat}{Nature}
\newcommand{\mnras}{MNRAS}

\newcommand{\pasj}{PASJ}
\newcommand{\apss}{APSS}
\newcommand{\pasp}{PASP}
\newcommand{\apj}{ApJ}
\newcommand{\aap}{A\&A}
\newcommand{\aaps}{A\&AS}

\newcommand{\actaa}{Acta Astronomica}

\newcommand{\aj}{AJ}
\newcommand{\apjl}{ApJ}

\newcommand{\sxp}{SXP\,1062}

\newcommand{\chandra}{{\em Chandra}}
\newcommand{\xmm}{{\em XMM-Newton}}
\newcommand{\swift}{{\em Swift}}
\newcommand{\salt}{{\em SALT}}

\def \change{}
\def \nchange{}

\begin{document}
\parindent=0cm


\maketitle

\label{firstpage}

\begin{abstract}

SXP\,1062 is a Be X-ray binary located in the Small Magellanic Cloud. It
hosts a long-period X-ray pulsar and is likely associated with the supernova
remnant MCSNR\,J0127-7332. In this work we present a multi-wavelength view 
on \sxp\ in different luminosity regimes. We consider monitoring 
campaigns in optical (OGLE survey) and X-ray (\swift\ telescope).    
During these campaigns a tight coincidence of X-ray  and  optical outbursts 
is observed. We interpret this as typical Type I outbursts as often 
detected in Be X-ray binaries at periastron passage of the neutron star. To 
study different X-ray luminosity regimes in depth, during the source quiescence 
we observed it with  \xmm\  while \chandra\ observations followed an X-ray 
outburst. Nearly simultaneously with \chandra\ observations in X-rays, in 
optical the  RSS/\salt\ telescope obtained spectra of \sxp. 
On the basis of our multi-wavelength campaign we propose a simple scenario 
where the disc of the Be star is observed face-on, while the orbit of the 
neutron star is inclined with respect to the disc.  According to the model of
quasi-spherical settling accretion our estimation of the magnetic field of the
pulsar in \sxp\ does not require an extremely strong magnetic field at the
present time.


\end{abstract}


\section{Introduction} 
\label{sec:intro}

Be X-ray binaries \citep[BeXBs;][]{Reig2011} are a subgroup of high mass X-ray 
binaries. The donor stars in these interesting objects 
are non-supergiant, fast-rotating B-type stars with {\change circum-stellar} 
decretion discs. 
The compact objects, in most confirmed BeXBs, are magnetized rotating 
neutron stars (NSs). Therefore, the majority of BeXBs are X-ray pulsars 
\citep[][]{Nagase1989}. The accretion in BeXBs is likely powered by  
the interaction between the decretion disc of the Be star and the NS 
\citep[][]{Okazaki2001}. 

The subject of this study is an X-ray pulsar discovered with \chandra\ and 
\xmm\ in March 2010 in the Wing of the Small Magellanic Cloud (SMC) galaxy 
\citep{Henault2012,Haberl2012}.  From the analysis of these first 
observations a long X-ray pulse period  $P_{\rm spin}\sim1062$\,s was found, 
accordingly the pulsar was named \sxp. The optical counterpart of \sxp\  is the 
Be star 2dFS\,3831 \citep[B0-0.5III;][]{Evans2004}.

\begin{table*}
\begin{center}
\caption{Log of X-ray \chandra\ and \xmm\ observations analyzed in 
this paper.  
\label{tab:xrayobservations}}
\begin{tabular}{ccccccc}
\hline
\# & Instrument & ObsID &  MJD & Date & Exp Time (ks) \\
\hline
1 & EPIC-MOS1/\xmm & 0721960101 &  56606.79 & 2013-10-11 & 77.1  \\
1 & EPIC-MOS2/\xmm & 0721960101 &  56606.79 & 2013-10-11 & 77.2  \\
1 & EPIC-pn/\xmm   & 0721960101 &  56606.80 & 2013-10-11 & 66.2  \\
2 & ACIS/\chandra  & 15784      &  56827.80 & 2014-06-19 & 29.6  \\
3 & ACIS/\chandra  & 15785      &  56837.49 & 2014-06-29 & 26.8  \\
4 & ACIS/\chandra  & 15786      &  56846.30 & 2014-07-08 & 26.3  \\
5 & ACIS/\chandra  & 15787      &  56856.91 & 2014-07-18 & 25.0  \\
\hline
\end{tabular}
\end{center}
{ NOTE:} The exposure time is the sum of 
good time intervals after removing times of high background.
\end{table*}

\sxp\ is especially interesting because it is the first X-ray pulsar 
discovered within its likely parental supernova remnant (SNR).  The association 
of SXP\,1062 with 
MCSNR J0127-7332\footnote{\url{http://sci.esa.int/xmm-newton/49789}} allows 
to estimate the age of the NS. Two 
independent groups obtained compatible values for the kinematic age of 
MCSNR J0127-7332 of $\sim$2$-$4$\times10^4$\,years \citep{Henault2012}, and 
$\sim$1.6$\times10^4$\,years \citep{Haberl2012}. 
According to standard models, at this age the NSs are 
too young to be in the accretor phase, and, consequently,  are not expected 
to be  X-ray pulsars \citep[][and references therein]{lipunovbook1992}. 

A number of models and scenarios are suggested to explain the 
physics of \sxp.  \cite{Haberl2012} proposed that its NS was born 
with unusually long initial spin period significantly larger than $0.1$\,s. 
\cite{Popov2012sxp} suggested that the NS is spinning close to the 
equilibrium period and has a magnetic field of 
$B\lesssim10^{13}$\,G at present time, while initially it had a strong 
magnetic field in excess of $B\sim10^{14}$\,G. \cite{Ikhsanov2012} put 
forward  a magnetic accretion scenario, while \cite{Fu2012} explored the 
possibility that \sxp\ is a present day accreting magnetar.

In this paper we use a complex multi-wavelength approach and analyse a large 
set of data with the goal to shed new light on the origin and properties of 
\sxp. The data were collected during the quiescent stage as well as at
outburst. The optical observations were obtained with the \salt\ 
telescope; we also use publicly available OGLE data. The X-ray observations 
analysed in this paper are obtained with \swift\ (long time monitoring), 
\xmm\ (in quiescence) and \chandra\ (in outburst). 
The optical and X-ray observations were coordinated  during 
outburst. The paper is organized as follows. 
Section\,\ref{sec:obs} presents the new data and observations. 
Section\,\ref{sec:analysis} is dedicated to the data analysis and results. In 
Section\,\ref{sec:dis} we compare our results with previous results and discuss 
the enigmatic properties of \sxp. 
Finally, in Section\,\ref{sec:conclusion} we present a summary  and 
our final conclusions.


\section{Observations}
\label{sec:obs}

\subsection{X-ray data}
\label{sec:x}

\begin{figure*}
\begin{center}
\includegraphics[angle=0,width=0.9\linewidth]{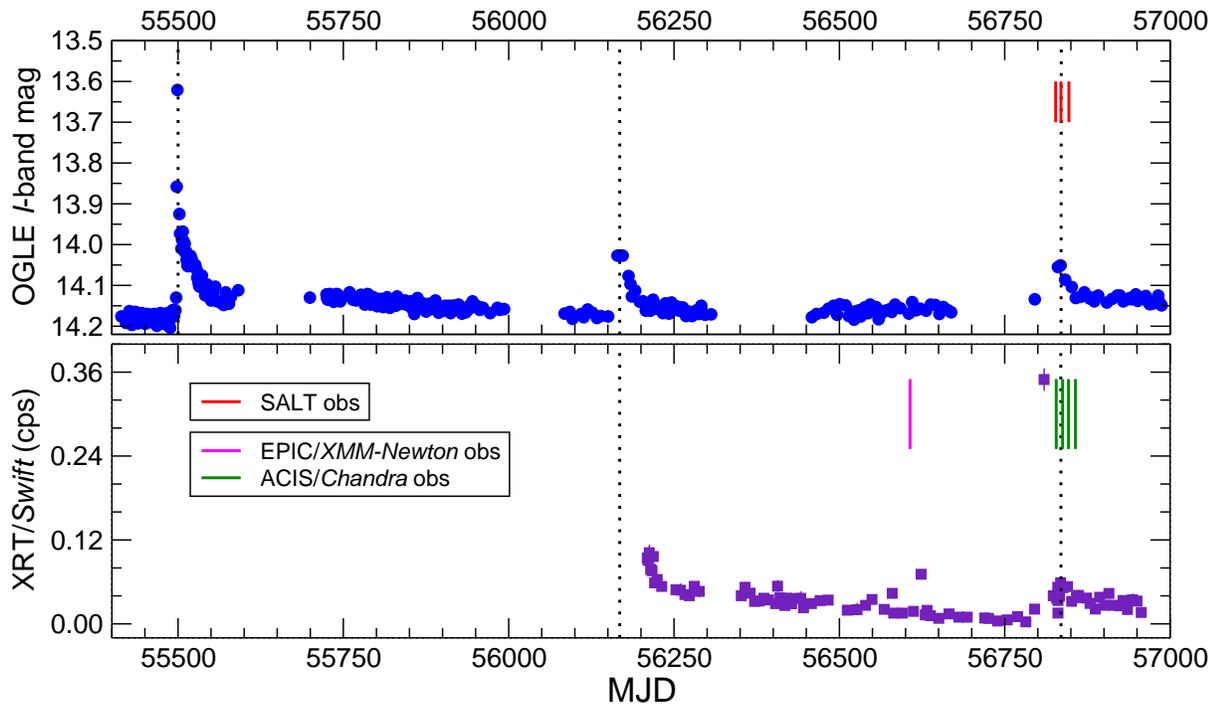}
\caption{{\it Upper panel:} OGLE {\it I}-band light curve covering \sxp\ between 
2010-08-06 (MJD\,55414.14) and 2014-11-26 (MJD\,56987.10). The red vertical 
lines indicate the epochs of the optical RSS/\salt\, observations. {\it Bottom 
panel:} XRT/Swift $0.3-10$\,keV X-ray light curve between 2012-10-09 
(MJD\,56209) and 2014-06-22 (MJD\,56990). The pink vertical line 
indicates 
the epoch of the EPIC observation, and the green vertical lines indicate 
the epochs of the ACIS observations. The dotted vertical lines shown 
in both panels indicate the periastron passage of NS according to our analysis 
(Section\,\ref{subsec:orbitalper}). Note the coincidence between the X-ray 
outburst shown in the bottom panel with the third optical outburst shown in the 
upper panel. \label{fig:outbursts}}
\end{center}
\end{figure*}

The X-ray Telescope (XRT) mounted on the \swift\ observatory \citep{Gehrels2004}
operates in the energy range $0.3-10$\,keV. XRT was monitoring \sxp\ between 
2012-10-09 and 2014-10-26. Figure\,\ref{fig:outbursts} displays the 
\swift\ X-ray lightcurve of \sxp.

During quiescence of \sxp\ (pink vertical line in the bottom panel of 
Fig.\,\ref{fig:outbursts}) we obtained an \xmm\ observation with the  
European Photon Imaging Camera \citep[EPIC;][]{Turner2001,Struder2001} 
(Table\,\ref{tab:xrayobservations}). EPIC comprises a 
set of three X-ray CCD cameras: EPIC-MOS1, EPIC-MOS2, and EPIC-pn. 

During the \swift\ monitoring campaign an X-ray outburst was detected on 
2014-06-01. The detection of the X-ray outburst triggered our target of 
opportunity \chandra\ observations with the Advanced CCD Imaging 
Spectrometer \citep[ACIS;][]{Garmire2001}. We obtained four \chandra\
observations (see Table\,\ref{tab:xrayobservations}). Dates of 
observations are marked in Fig.\,\ref{fig:outbursts} as green vertical 
lines. The first \chandra\ observation 
was performed using the {\it Continuous Clocking Mode} which is recommended for 
very bright sources  as it prevents the photon pile-up.  The other three 
exposures were obtained with the standard {\it Timed} read out mode. 

\subsection{Optical data}

\subsubsection{RSS/SALT}
\label{sec:obssalt}

Optical observations were obtained by the 11\,m telescope \salt\ telescope 
nearly simultaneously with the \chandra\ observations during \sxp\ outburst.
We obtained four sets of observations (see 
Table\,\ref{tab:opticalobservations}). The epochs of three data 
sets obtained in 2014 are 
indicated in the top-panel of Fig.\,\ref{fig:outbursts} as red 
vertical lines.

The optical spectroscopy was collected using the Robert 
Stobie  dual-beam Visible/Near-IR spectrograph 
telescope \citep[RSS;][]{Kobulnicky2003}.
The data sets consist of long-slit spectra  using two 
different long-slit set-ups. The first covers wavelengths between $6082$\,\AA\ and 
$6925$\,\AA, i.e., the H$\alpha$ emission line, with a slit width of either 
$1.25''$ (first observation) or $1.5''$ (rest of observations). The second 
set-up covers the range between $3835$\,\AA\ and $4924$\,\AA, i.e., H$\beta$ and 
H$\gamma$ emission lines with a slit width of $1.5''$.

\begin{table}
\begin{center}
\caption{Equivalent widths of the H$\alpha$ line in the optical spectra 
of \sxp. \label{tab:halpha}}
\begin{tabular}{ccc}
\hline
MJD & $EW$ (\AA) &   Reference \\
    & H$\alpha$  & \\
\hline
55494 & $-23 \pm 1$     & \cite{Henault2012}\\
56214 & $-26.7 \pm 0.1$ & \cite{Sturm2013}\\
56827 & $-36.3 \pm 0.6$ & This work \\
56835 & $-34.0 \pm 0.6$ & This work \\
56847 & $-34.1 \pm 0.6$ & This work \\
57694 & $-33 \pm 1$     & This work \\
\hline
\end{tabular}
\end{center}
\end{table}

The data were first processed using the 
{\small PySALT}\footnote{\url{http://pysalt.salt.ac.za}} package. 
Subsequent flat-fielding, background subtraction, wavelength calibration, and 
extraction of $1D$ spectra were performed with the Image Reduction and Analysis 
Facility ({\small IRAF}\footnote{\url{http://iraf.noao.edu}}). 
After wavelength calibration and 
extraction, the final spectra were red-shift corrected for the systemic velocity 
of the SMC (150\,km\,s$^{-1}$).

The equivalent width ($EW$) calculated by fitting a double Gaussian 
model are summarized in Table\,\ref{tab:halpha} together with the previously 
published values for comparison. 

\begin{figure*}
\begin{center}
\includegraphics[width=0.85\linewidth]{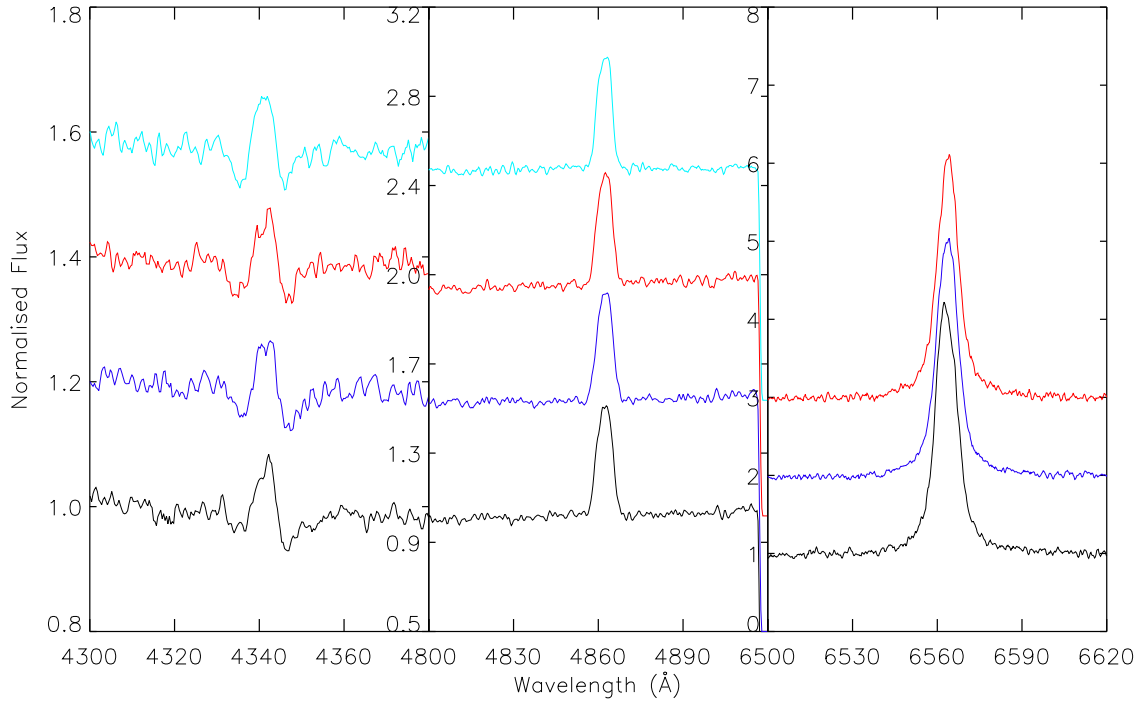}
\caption{\sxp\ H$\gamma$ (left), H$\beta$ (middle) and H$\alpha$ (right) 
profiles
as observed during June 2014 after the outburst detected with \swift. 
Three different H$\alpha$ profiles were obtained during three different 
nights, and four H$\beta$ and H$\gamma$ profiles observed during two different 
nights (two spectra per night) as summarized 
in Table\,\ref{tab:opticalobservations}. The spectra are displayed with an 
arbitrary offset for presentation purposes and are ordered by date increasing 
from bottom to top. 
\label{fig:halpha}}
\end{center}
\end{figure*}

Figure\,\ref{fig:halpha} displays the observed H$\gamma$, H$\beta$ and 
H$\alpha$ profiles {\change obtained from the} RSS/\salt\ post-outburst 
observations. The strong H$\beta$ and H$\alpha$ lines show non-split symmetric 
shapes. The spectrum around the H$\gamma$ line is more noisy and the line 
structures apparent in some observed spectra are likely due to the noise. 

\subsubsection{OGLE}

The Optical Gravitational Lensing Experiment (OGLE) Monitoring 
system provides  optical {\it I}-band photometry of X-ray 
sources\footnote{\url{http://ogle.astrouw.edu.pl/ogle4/xrom/xrom.html}} located 
in the fields observed  with roughly daily sampling  by the OGLE-IV Survey 
\citep{Udalski2008}. To date there are seven seasons of data available for 
\sxp\ covering the time interval from MJD\,55414.4 (2010-08-06) to MJD\,57773.039 (2017-01-20).

\section{Data Analysis \& Results}
\label{sec:analysis}

\subsection{Determination of orbital period based on the optical and X-ray 
light curves}
\label{subsec:orbitalper}

In general, Be stars show photometric variability on various time scales 
\citep{Porter2003}. Stellar rotation and pulsations are held responsible  
for fast variability, while long-term variability on time scales of a year 
is typical for formation and dissipation of the disc \citep{Jones2008}.
{\change \citet{Reig2015}  point out that since the disc constitutes the main 
reservoir of matter available for accretion,  a correlation between X-ray and 
optical emission should be expected. This is indeed often observed 
\citep[e.g.][]{coe2009}. In particular, a few sources are known where the 
minimum magnitude in optical coincides with the X-ray outburst \citep[e.g., 
SXP\,46.6, see Fig. 4 in][]{McGowan2008}. Importantly, the amplitude of 
the orbital variation $\sim 0.6$\,mag observed in \sxp\ is very 
similar to those observed in other systems, such as e.g.\ AX\,J0049.4-7323, 
\citep{Cowley2003}.

Among possible explanations of optical outbursts is the tidal interaction of 
the neutron star with the Be disk. Such interactions lead to
strong deformations of the disk during periastron passage  \citep{Okazaki2001}.
As discussed by \citet{McGowan2008} and \citet{coe2009}, this might
lead to an increase in disk size, which then translates to an increase in
optical brightness. Even spectral changes in the optical related to 
the brightness in the spectral energy distribution of BeXB systems have been 
observed in several systems. This can be explained by changes in the disk size 
which alters the contribution to the spectral energy distribution at longer 
wavelength \citep[see the discussion in][and references 
therein]{Vasil2014}. {\nchange However, the physical situation is likely more 
complex because in some BeXBs, e.g.\ A0535+262, the opposite behavior is 
observed -  during the X-ray outburst, the optical and IR fluxes are reduced 
\citep{Camero2012, Naik2012}.
}

No detailed numeric simulations of NS and Be-disk capable to predict 
the optical and X-ray light curves of individual systems exist yet. However, 
the coupling between the disc material and X-rays powered by accretion is 
strongly supported by observed coincidence between optical and X-ray 
variability.


While not all BeXBs show X-ray outbursts at every periastron passage,
on the basis of the previous significant work outlined above, we suggest 
that the time interval between outbursts in \sxp\ is the binary orbital 
period. Future observations constraining binary parameters are required to 
firmly confirm or disprove this suggestion.}

As can be seen in the upper panel in Fig.\,\ref{fig:outbursts}, \sxp\ 
shows regular outbursts in the optical. \cite{Schmidtke2012} proposed that the 
time interval of $656\pm2$\,d between the first two optical outbursts  is the 
orbital period of \sxp. \cite{Sturm2013} noticed the fast rise exponential 
decay (FRED)-shaped profile \citep{Bird2012} of these optical outbursts, which  
supports their orbital origin, i.e., the outbursts occur at periastron passage of the 
NS.

The current {\small OGLE-IV} {\it I}-band light curve covers more than 6\,yr.  
Three equi-distant outbursts were observed confirming their periodic nature and 
allowing to better constrain the orbital period.  Unfortunately, during the 
latest outburst predicted for 2016-04-26 (MJD 57504), there was a gap in the 
OGLE observations. Hence, the latest predicted outburst was  not recorded. 

To further tighten the orbital 
period, we selected the minimum magnitude detected in the optical OGLE {\it 
I}-band light curve as the mid-time for the outbursts. The 
spacing of the OGLE observations is somewhat sparse (roughly once a day), 
therefore we 
cannot precisely determine the exact time of the start of the outbursts. To overcome 
this difficulty, we took the time interval between the data points immediately 
before and after the mid-time as the uncertainty (second column of 
Table\,\ref{tab:orbitalperiod}). As a next step,  we calculated the time 
intervals between the outbursts (third column of 
Table\,\ref{tab:orbitalperiod}), and the weighted average of these time 
intervals. 

{\nchange One shall be aware, that in some rare cases,  such as 
XTE\,J1946+274, the X-ray outbursts might not be locked with 
an orbital period, or happen more than once per orbit 
\citep{Wilson2003}. Nevertheless, the simplest assumption is that 
X-ray outbursts in \sxp\ occur during the periastron passage. } 
As can be seen in Fig.\,\ref{fig:outbursts}, the X-ray outbursts from \sxp\
are regular and happen nearly at the same time as the optical ones. 
During our \swift\ monitoring campaign we have detected an X-ray outburst in
excellent agreement with the predictions. This gives further credibility to 
the scenario where X-ray and optical outbursts in \sxp\ occur simultaneously at 
the periastron passage of the NS. 
Figure\,\ref{fig:outbursts} also suggests that the minimum optical magnitude 
of each outburst is apparently increasing with time, i.e. $\Delta$I is 
decreasing. This could be either of physical origin or due to the sparse data sampling.

\begin{table}
\begin{center}
\caption{Outburst history of \sxp.
\label{tab:orbitalperiod}}
\begin{tabular}{cccc}
\hline
Outburst & MJD &  $\Delta$t (days) & $\Delta$I (mag)\\
\hline
1 & $55500.09\pm0.97$ & & $\gtrsim0.55$ \\
2 & $56166\pm14$& $666\pm7$  & $\gtrsim0.15$ \\
3 & $56835.9\pm4.5$ & $670\pm8$ & $\gtrsim0.11$ \\
\hline
\end{tabular}
\end{center}
{ NOTE:} The first column indicates the outburst ordered by date, the second 
column
the outburst time. The third column provides the time interval between 
outbursts and the fourth column the amplitude of the outburst. 
\end{table}

\subsection{X-ray spectrum in quiescence measured by EPIC/XMM-Newton }
\label{sec:xq}

\begin{figure}
\begin{center}
\includegraphics[width=0.7\columnwidth, angle=-90]{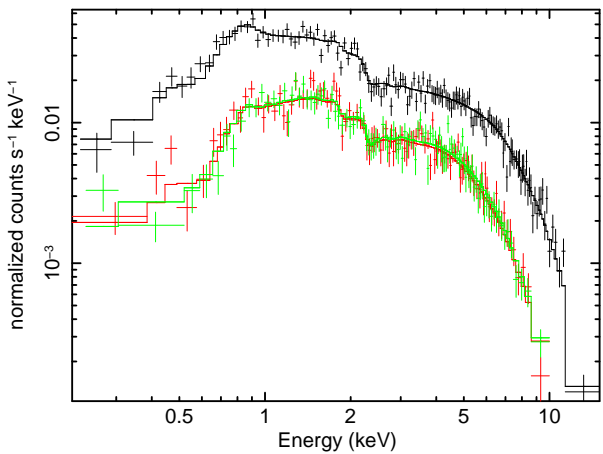}
\caption{EPIC X-ray spectra of \sxp\ in quiescence as observed on 2013-10-11 
together with the best-fit model 
(solid lines, see Table\,\ref{tab:specfit} for model parameters). 
EPIC-pn/MOS1/MOS2 data are plotted in black, red and green 
correspondingly. 
\label{fig:xmmspec}}
\end{center}
\end{figure}

\begin{figure*}
\begin{center}
\includegraphics[width=0.7\linewidth]{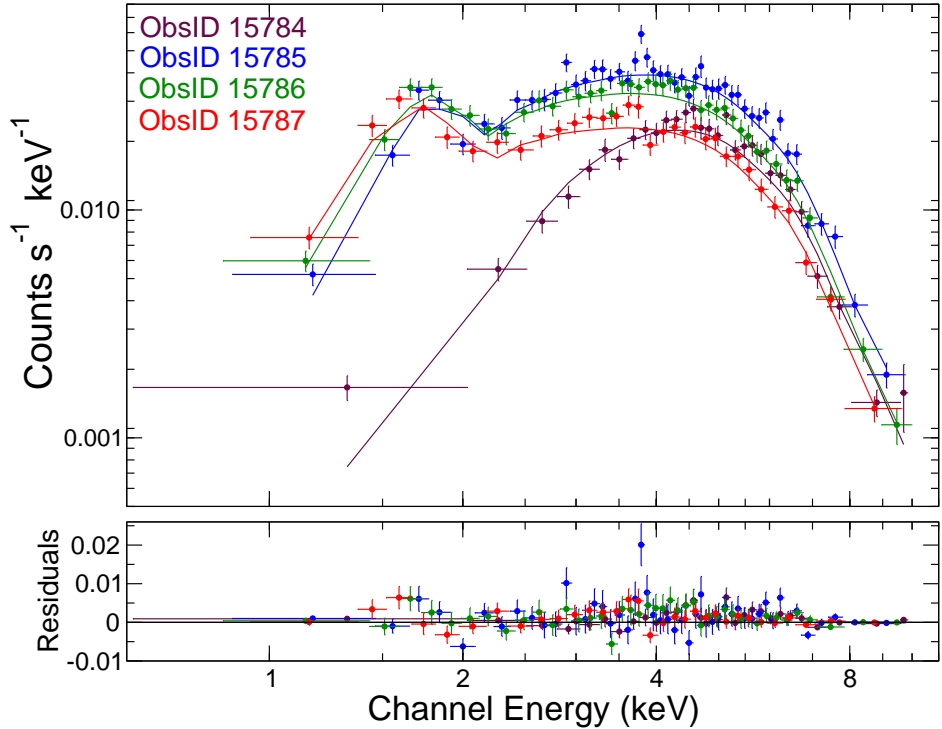}
\caption{ACIS post-outburst X-ray spectra of \sxp\ as observed on 2014-06-19
(brown), 2014-06-29 (blue), 2014-07-08 (green), and 2017-07-18 (red). 
The upper panel shows the spectra (dots 
with error bars) together with the best-fit model (solid line), while the lower 
panel shows the residuals of the model with respect to the 
spectra.\label{fig:chandraspec}}
\end{center}
\end{figure*}

\begin{table*}
\begin{center}
\caption{Previously published absorbed X-ray fluxes and un-absorbed X-ray 
luminosities for \sxp\ 
together with measurements obtained during this work. \label{tab:lum}}
\begin{tabular}{cccccc}
\hline
Date & MJD & $\Delta E$ & $F_{\rm X}$\,(erg\,cm$^{-2}$\,s$^{-1}$) & $L_{\rm 
X}$\,(erg\,s$^{-1}$) & Reference \\
\hline
2010-03-25 & 55280 & $0.2-12$\,keV & 
$\left(1.7^{+0.2}_{-0.3}\right)\times10^{-12}$ & 
$\left(6.3^{+0.7}_{-0.8}\right)\times10^{35}$ & \cite{Henault2012} \\ [0.2cm]
2012-10-14 & 56214 & $0.2-10$\,keV & $\left(6.9\pm0.2\right)\times10^{-12}$ & 
$2.6\times 10^{36}$ & \cite{Sturm2013} \\ [0.2cm]
2013-10-11 & 56606.80 & $0.2-12$\,keV & 
$\left(1.39^{+0.01}_{-0.01}\right)\times10^{-12}$ & 
$\left(5.4\pm0.1\right)\times10^{35}$ & This work \\ [0.2cm]
2014-06-19 & 56827.80 & $0.3-10$\,keV & 
$\left(6.2^{+0.1}_{-0.5}\right)\times10^{-12}$ & 
$\left(2.6\pm0.1\right)\times10^{36}$ & This work \\ [0.2cm]
2014-06-29 & 56837.49 & $0.3-10$\,keV & 
$\left(7.6^{+0.1}_{-0.2}\right)\times10^{-12}$ & 
$\left(3.0\pm{0.1}\right)\times10^{36}$ & This work \\ [0.2cm]
2014-07-08 & 56849.30 & $0.3-10$\,keV & 
$\left(6.0^{+0.1}_{-0.2}\right)\times10^{-12}$ & 
$\left(2.4\pm0.1\right)\times10^{36}$ & This work \\ [0.2cm]
2014-07-18 & 56856.91 & $0.3-10$\,keV & 
$\left(4.2^{+0.1}_{-0.1}\right)\times10^{-12}$ & 
$\left(1.6\pm0.1\right)\times10^{36}$ & This work \\
\hline
\end{tabular}
\end{center}
\begin{flushleft}
{ NOTE:} X-ray luminosities are scaled to $DM=18.7$. The 
X-ray absorbed flux 
errors of this work are given for a 68\% confidence level, while the 
un-absorbed X-ray luminosities are given for a 90\% confidence level.
\end{flushleft}
\end{table*}


X-ray spectra during outburst and quiescence were obtained by \chandra\ 
and \xmm, respectively (see section \ref{sec:x}). The spectral analysis
was performed using the X-ray spectral fitting package {\small XSPEC} 
\citep{Arnaud1996}. All  spectra were fitted using a model which includes 
three absorption components. 
The first component accounts for the absorption in the Milky Way with a  
fixed column density $6\times10^{20}$\,cm$^{-2}$ and solar abundances 
\citep{Wilms2000}. The second component 
accounts for the local absorption in the SMC, and has a fixed column density  
$2\times10^{21}$\,cm$^{-2}$ \citep{Oskinova2013} and SMC 
abundances \citep[0.2 times solar for elements heavier than 
He;][]{Russell1992}. The third component, also with SMC 
abundances, accounts for the absorption intrinsic to \sxp. We keep this 
absorption component as a free parameter during the spectral fitting 
procedure. 

X-ray luminosities in the energy range 0.2-15\,keV assuming a 
distance modulus to the SMC Wing of $DM=18.7$ \citep{Cignoni2009} as well 
as X-ray fluxes  are given in Tables\,\ref{tab:lum}. 

We start by considering the X-ray spectra measured during the quiescence by 
\xmm\  (see Fig.\,\ref{fig:xmmspec}). As a first step we have simultaneously 
fitted all three EPIC spectra (pn, MOS1 and MOS2) with an absorbed 
power law. The model fit shows residuals in the soft part 
of the spectra ({\nchange $\chi^2= 1.36$}). Therefore, as a next step we tried 
to refine the model adding a 
black-body component. This led to a statistical improvement of the 
fit ({\nchange $\chi^2= 1.26$}). We also tested a combination of a power law 
and an optically thin thermal 
plasma component \citep[APEC;][]{Smith2001} with SMC abundances. However, 
including this component does not improve the fit. Finally, the best fit 
is obtained using a power law together with a black-body component, and the 
thermal plasma component ({\nchange $\chi^2= 1.05$}). This best-fit model is 
shown in Fig.\,\ref{fig:xmmspec}, 
and the corresponding parameters for the various models are given in 
Table\,\ref{tab:specfit}. Similar X-ray spectral properties of \sxp\ 
were 
previously derived by \citet{Henault2012} from the analysis of observations 
obtained in March 2010.  

\begin{table*}
\begin{center}
\caption{Spectral models of \sxp\ in quiescence and post-outburst (see text for 
details).
\label{tab:specfit}}
\begin{tabular}{cccccc}
\hline
MJD & $N_{\rm H}$            & $\Gamma$ & $kT_{\rm BB}$ & $kT_{\rm Thermal}$ 
& $\chi^2$ \\
    & ($10^{22}$\,cm$^{-2}$) &          &   [keV]       &  [keV]  &   \\
\hline 
\multicolumn{6}{c}{\xmm\ observations during quescence} \\
\hline
56606.79 & $0.5^{+0.2}_{-0.1}$ & $0.8^{+0.2}_{-0.1}$ &  $1.7^{+0.2}_{-0.2}$ &  
$0.8\pm0.1$ & 1.05 \\
\hline 
\multicolumn{6}{c}{\chandra\ observations following outburst} \\
\hline 
56827.80 & $177^{+38}_{-31}$    & $-0.1\pm 0.2$& & & 1.22  \\ [0.2cm]
56837.49 & $42^{+5}_{-5}$       & $0.4\pm 0.1$ & & & 1.10  \\ [0.2cm]
56846.30 & $31.6^{+3.9}_{-3.5}$ & $0.5\pm 0.1$ & & & 1.09  \\ [0.2cm]
56856.91 & $20.2^{+3.6}_{-3.2}$ & $0.5\pm0.1$  & & & 0.99  \\ 
\hline
\end{tabular}
\end{center}
\begin{flushleft}
{ NOTE:} Errors are given for a confidence level of 90\%.
\end{flushleft}
\end{table*}

A  black-body spectral component characterized by $kT\gtrsim1.5$\,keV seems to 
be a common feature of X-ray spectra of low luminosity ($L_{\rm 
X}\lesssim10^{36}$\,erg\,s$^{-1}$) and long spin period 
($P_{\rm spin}\gtrsim100$\,s) X-ray pulsars \citep[e.g.][]{LaPalombara2013}. 
This component contributes 30-40\% of the source flux below 10\,keV and 
implies a small emission radius ($R_{\rm BB}<0.5$\,km) which is typically 
attributed to the emission from the NS polar caps with a size of 
$\sim0.1R_{\rm NS}$ \citep{Hickox2004}. 

The X-ray luminosity of the black-body component in the
X-ray spectrum of \sxp\ in quiescence, $L_{\rm X}\approx \left(1.8\pm0.1\right)\times 
10^{35}$\,erg\,s$^{-1}$, 
amounts to  $\sim$40\% of the total X-ray emission below $10$\,keV, 
similar to other long-period pulsars. Using the Stefan-Boltzmann law, 
and the temperature given in Table\,\ref{tab:specfit}, 
we estimate the size of the polar cap emission region in \sxp\
to $R_{\rm BB}=400^{+140}_{-90}$\,m.  

{\change Our analysis shows that in \sxp\ the thermal plasma spectral 
component} accounts for $\sim$20\% of the 
X-ray flux in the softest energy band ($0.2-1.0$\,keV). 
Massive stars with spectral types earlier than B2 are sources of thermal 
X-ray emission \citep[e.g.,][]{Berghoefer1997}. The typical X-ray luminosity 
of a B0III star does not exceed $\sim$10$^{32}$\,erg\,s$^{-1}$ and the bulk of 
this emission is quite soft with temperatures of about 
10$^6$\,K \citep{Raassen2005, Fossati2015}. Yet, the luminosity of coronal 
plasma in \sxp\ is nearly one order of magnitude larger, 
$L_{\rm X}\approx 6\times10^{33}$\,erg\,s$^{-1}$.
Hence, the intrinsic X-ray emission from the stellar wind of the B0IIIe-type 
donor star cannot fully account for the thermal radiation observed from this source.
We speculate that this soft component originates, at least partly, in the 
diffuse 
emission from the maternal SNR. 

\subsection{Post-outburst X-ray spectra measured by \chandra}
\label{sec:cho}

\chandra\ obtained four spectra following the outburst detected by \swift\
(see Fig.\,\ref{fig:outbursts}). The first post-outburst observation 
(ObsID\,15784) has significantly lower X-ray flux than the following observation 
(ObsID\,15785). This sharp post-outburst drop in X-ray flux and its subsequent 
recovery is also evident in the  XRT/\swift\ light curve. As discussed later  
in Section\,\ref{sec:disc}, the flux decrease is mainly due to the strong 
increase in the absorbing column.  

To model the \chandra\ spectra we started with fitting a simple absorbed power law  
(see Fig.\,\ref{fig:chandraspec} and Table\,\ref{tab:specfit}). 
Subsequently, by analogy with the quiescence spectra we attempted to include 
a thermal plasma component, but the fits did not improve. 

Iron K$\alpha$ emission is often observed in persistent high-mass X-ray 
binaries \citep[e.g.,][]{Gimenez2015} and in 
transient BeXBs during outbursts \citep{Reig2011} due to the increase of 
matter density
surrounding the NS. \cite{Sturm2013} found an indication for the presence of 
an Fe-K$\alpha$ line in their \xmm\ spectra of \sxp\ during an outburst in 
2012. We also tested for the presence of the iron fluorescence emission line at 
6.4\,keV in our \chandra\ spectra but this did not lead to a significant improvement
of the fit.
This non-detection might be due to a poor signal to noise ratio in our 
ACIS spectra compared to previous \xmm\ observations.

\subsection{X-ray pulse periods and profiles measured by \xmm\ and \chandra}
\label{sec:pulseperiods}

\begin{figure}
\begin{center}
\includegraphics[angle=0,width=1.0\linewidth]{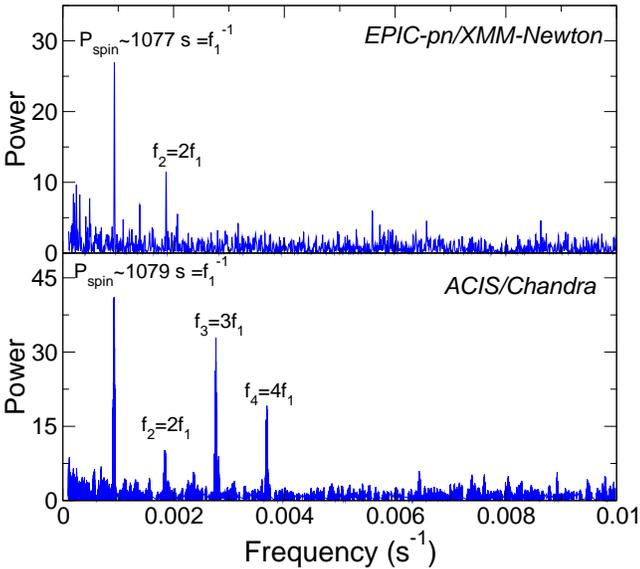}
\caption{Lomb-Scargle periodograms of EPIC-pn (MJD\,56606.80) 
and merged ACIS (MJDs 56827.80, 
56837.49, 56846.30, and 56856.91) observations of \sxp\ obtained with the 
{\small PERIOD} package within the {\it Starlink} environment. The main peak of 
the EPIC-pn periodogram corresponds to $P_{\rm spin}\sim1077$\,s, and the 
main peak of the merged ACIS light curve corresponds to $P_{\rm 
spin}\sim1079$\,s.  In both cases peaks at harmonic frequencies are seen. 
A significance calculation with 1000 permutations of the data 
through a Fisher randomization was performed through the {\small SIG} option of 
{\small PERIOD} obtaining a significance level of $\gtrsim$95\% for the highest 
peak in both periodograms. \label{fig:scargle}}
\end{center}
\end{figure}

\begin{figure}
\begin{center}
\includegraphics[angle=0,width=1.0\linewidth]{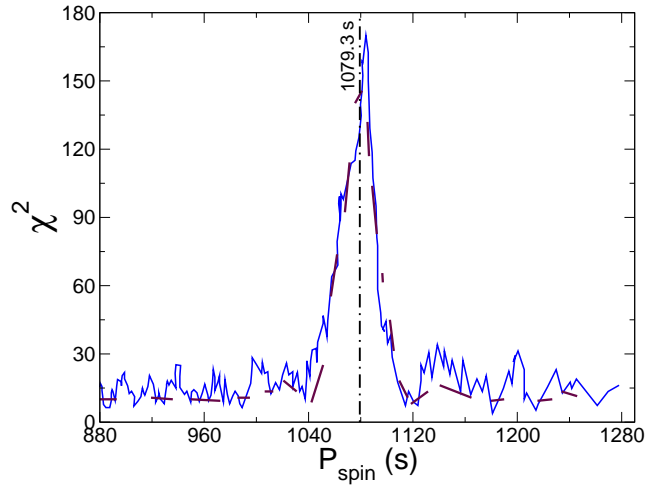}
\caption{Example of an epoch folding analysis of the ACIS light curve of \sxp\ 
from 2014-07-08, performed with the 
software package DES7. The continuous line is the $\chi^2$ distribution that corresponds 
to the light curve, the dashed line indicates the $\chi^2$ 
template function, and the vertical line marks the spin period derived from this 
observation (for more details see Section\,\ref{sec:xq}). 
\label{fig:chi2}}
\end{center}
\end{figure}

\begin{figure}
\begin{center}
\includegraphics[angle=0,clip=,width=0.8\linewidth]{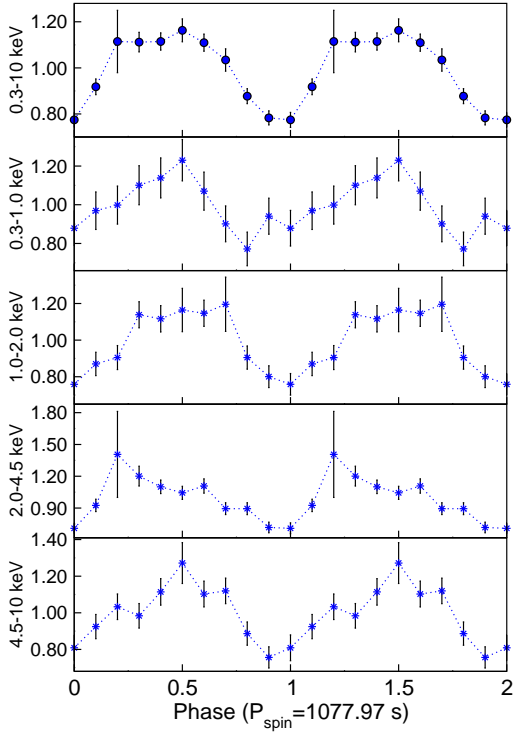}
\caption{Pulse profiles obtained from EPIC-pn light curves folded over 
$P_{\rm spin}=1077.97$\,s as observed on 2013-10-11 in quiescence. The 
top panel shows the pulse profile from the 0.3$-$10\,keV energy band, while in the 
other panels pulse profiles for different sub-energy bands are presented. 
Every single light curve is background subtracted and normalised to the average 
count rate. Phase zero corresponds to the phase of minimum normalized intensity 
of the folded light curve from the broad energy band. Pulse 
profiles are plotted twice for clarity.\label{fig:pulsexmm}}
\end{center}
\end{figure}

\begin{table*}
\begin{center}
\caption{Spin, $P_{\rm spin}$, and spin derivative, $P_{\rm spin}$,  
measurements of \sxp\ \label{tab:spin}}
\begin{tabular}{cccc|ccc}
\hline
Date       & MJD      & Telescope & $P_{\rm spin}$ & $\Delta t$ &  $\dot{P}_{\rm 
spin}$ & Confidence \\
           &          &           & [s]            &   [d]      &   
[s\,d$^{-1}$]       &                       \\ 
\hline
2010-03-25 & 55280    & \xmm\     & $1062$              & 18  & $0.24\pm0.07$    
  & 99.999\% \\
2012-10-14 & 56214    & \xmm\     & $1071.01\pm0.16$    & 915 & 
$0.0062\pm0.0012$  & 43.658\% \\
2013-10-11 & 56606.80 & \xmm\     & $1077.97\pm0.28$ & 393 & $0.0177\pm0.0009$  
& 76.465\% \\
2014-06-19 & 56827.80 & \chandra\ & $1091.1\pm1.2$   & 221 & $0.059\pm0.006$    
& 95.982\% \\
2014-06-29 & 56837.49 & \chandra\ & $1087.1\pm0.9$   & 9   & $-0.45\pm0.16$     
& 99.999\% \\
2014-07-08 & 56846.30 & \chandra\ & $1079.3\pm1.9$   & 9   & $-0.86\pm0.23$     
& 99.999\% \\
2014-07-18 & 56856.91 & \chandra\ & $1086.0\pm4.4$   & 11  & $0.61\pm0.44$      
& 99.999\% \\
\hline
\end{tabular}
\end{center}
\begin{flushleft}
{ NOTE:} MJDs of the $\dot{P}$ values obtained during this work are the 
mid-time  value between two spin period measurements. The fifth  column, 
($\Delta$t), gives the time interval for which  $\dot{P}_{\rm spin}$ was 
calculated. The last column is the probability of the spin derivative to be 
intrinsic according to the probability distribution shown 
in Fig.\,\ref{fig:pdotorbit}. The parameters for the 2010-2012 epoch are from 
\cite{Henault2012, Haberl2012, Sturm2013}, otherwise as obtained in this work.
\end{flushleft}
\end{table*}

To study the pulsational behavior of \sxp\ we consider the energy bands where 
the signal from the source dominates over noise, i.e.\ 
$0.2-12$\,keV for EPIC and $1-8$\,keV for ACIS (see 
Figs.\,\ref{fig:xmmspec} and\,\ref{fig:chandraspec}).
EPIC-pn and ACIS light curves were extracted with a time binning of 
50\,s, corrected to the Solar System barycentre using the appropriate software 
routines, and background subtracted. Over a time interval of $\sim$29\,days 
and after removing high background intervals,  joined ACIS light curves
cover 107\,ks.


The  ACIS and the EPIC-pn light curves were 
searched for periodic signals between 100 and 10000\,s using the {\small 
SCARGLE} routine  of the {\it Starlink} 
software\footnote{\url{http://starlink.jach.hawaii.edu/starlink}}. This routine 
performs a fast implementation of the Lomb-Scargle periodogram using Fast 
Fourier Transforms to increase the speed of computation \citep{Press1989}. We 
performed a significance calculation by enabling the {\small SIG} option of 
{\small PERIOD} which runs a Fisher randomization test with a number of 
permutations set by the user (1000 in our case) at the same time as the period 
search is executed. This task gives as result two false alarm probabilities: 
{\it FAP1}, which is the probability that, given the frequency search 
parameters, 
there is no periodic component present in the data with this period; and 
{\it FAP2}, 
 which is the probability that the period is not actually equal to the quoted 
value but is equal to some other value. We obtained a period of $\sim$1079\,s 
(see bottom panel of Fig.\,\ref{fig:scargle}) with ${\it FAP1}=0$ and 
${\it FAP2}=0$ for the 
merged ACIS light curve, and a $P_{\rm spin}\sim1077$\,s (see top 
panel of Fig.\,\ref{fig:scargle}) with ${\it FAP1}=0$ and ${\it FAP2}=0$ for 
the 
EPIC-pn light curve. According to the {\small PERIOD} user manual, these 
{\it FAP} values imply that the false alarm probabilities lie between 0.00 and 
0.01 
with 95\% confidence. Both periodograms show several peaks (see 
Fig.\,\ref{fig:scargle}), the main peak is the spin period of \sxp\ at that 
time, while the other peaks correspond to harmonics of the fundamental frequency.

Period determinations for the EPIC-pn and the four ACIS 
light curves  were also performed with an epoch-folding analysis 
implemented in the {\small DES7} software package \citep[][]{Larsson1996}. 
This routine folds the data at a number of different test periods (this number 
was set to 200 periods in this analysis) around an initial input period. The 
input period is the output obtained with the Lomb-Scargle 
periodograms shown in Fig.\,\ref{fig:scargle}. For each period, the $\chi^2$ 
over the resulting pulse profile is computed, Fig.\,\ref{fig:chi2} shows an 
example of the distribution of $\chi^2$ vs. different test periods for one of our 
ACIS observations. A best-fit period is determined by fitting a 
$\chi^2$ template function to the observed $\chi^2$ distribution 
(Fig.\,\ref{fig:chi2} also shows this $\chi^2$ template). This $\chi^2$ template 
function is obtained by performing the same analysis to an artificial light 
curve that takes the time sampling of the data into account and, in an iterative 
procedure, the pulse shape of the oscillation. Uncertainties are estimated by 
Monte Carlo simulations. We have performed 500 Monte Carlo simulations in this 
analysis. A set of synthetic pulse light curves with the same time sampling and 
noise level as the data are created and analysed. The distribution of the 
periods determined during these simulations is used as a measurement of period 
uncertainty. The results of the analysis are shown in Table\,\ref{tab:spin} 
together with the previous published spin periods for \sxp.


To calculate the period derivatives (see Table\,\ref{tab:spin}) we 
assumed a constant spin down/up between every two subsequent $P_{\rm spin}$ 
measurements of the pulsar. This is a good approximation for 
\chandra\ given the short time interval between the individual 
measurements. However, the time interval between the EPIC-pn  
and the first ACIS measurement is too long with an X-ray outburst 
occurring in between. Therefore, this derivative can only be considered as an 
average value during that epoch, or even as a minimum value of the spin down if 
we consider that the pulsar started to spin up just after the outburst.

X-ray light curves were folded over the corresponding pulse period with the 
dedicated {\small XRONOS}\footnote{\url{
http://heasarc.gsfc.nasa.gov/docs/xanadu/xronos/xronos.html}} tool {\it efold}. 
This routine folds the light curve over a given period and normalizes it to the 
corresponding average count rate. Pulse profiles obtained from EPIC-pn 
data are shown in Fig.~\ref{fig:pulsexmm}. 

X-ray pulsars exhibit a wide variety of pulse shapes which differ from one 
source to another and which are strongly dependent not only on the energy band 
but also on  the X-ray luminosity. During quiescence the mass 
accretion rate is steady and low, the gas can freely fall onto the magnetic 
poles of the NS, and hence form a pencil beam emission. Therefore, X-ray 
pulsars 
in quiescence usually show smooth and single-peaked profiles. However, during 
periastron passage, the accretion of matter results in an increase of the 
X-ray flux, and consequently pulse profiles usually show the presence of dips 
at certain pulse phases which are more prominent at soft X-ray energies than 
at higher energies \citep[e.g., ][]{Naik2013}.

\begin{figure}
\begin{center}
\includegraphics[angle=0,clip=,width=0.75\columnwidth]{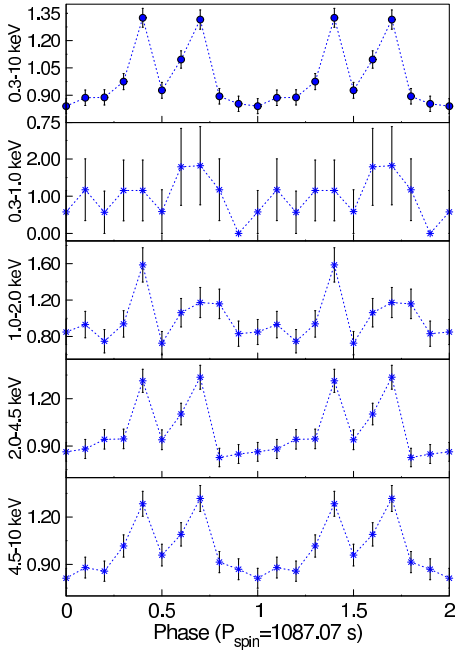}
\caption{\sxp\ pulse profile of ACIS post-outburst light curve folded 
over the corresponding spin period  as observed on
2014-06-29. The upper panel  
shows the pulse profile over the whole energy band, while the rest of the 
panels 
show pulse profiles for different sub-energy bands. Every single 
light curve is background subtracted and normalized to the  
corresponding average
count rate. Phase zero corresponds to the phase of minimum normalized intensity 
of the folded light curve that corresponds to the whole energy band. Pulse 
profiles are reported twice for clarity. \label{fig:pulsechandra}}
\end{center}
\end{figure}

Previous quiescence observations of \sxp\ in 2010 showed indications of a 
double peak in the pulse profile, while during the outburst observed in 2012 
there was only one clear peak at energies above 1\,keV \citep[see Fig.\,4 
of][]{Sturm2013}. This behavior is completely opposite to what is expected. 
However, during the quiescence observation reported in this paper we observe 
one smooth  peak (see Fig.\,\ref{fig:pulsexmm}), and there might be a small dip; 
while, during the \chandra\ post-outburst observations we can clearly see 
several dips (see Fig.\,\ref{fig:pulsechandra}); which is what we were 
expecting.

\subsection{Period derivative and orbital Doppler effect}
\label{sec:doppler}

The measured pulse period derivative of any X-ray pulsar is a combination of the 
intrinsic spin-up or spin-down and the Doppler shift caused by the orbital 
motion:

\begin{equation}
\label{eq:derivativeobs}
\dot{P}^{\rm obs}_{\rm spin}(t)=\frac{dP^{\rm obs}_{\rm spin}}{dt}=
\frac{dP_{\rm spin}(t)}{dt}+\frac{1}{c}\frac{d}{dt}\left(P_{\rm spin}(t)v_{\rm 
orb}(t)\right),
\end{equation}
where $v_{\rm orb}$ is the radial velocity, i.e. the projected orbital velocity along the line of sight:
\begin{equation}
\label{eq:projvel}
v_{\rm orb}(t)=\frac{2\pi a_{\rm X} \sin i}{P_{\rm orb} \sqrt{1-e^2}}\left(\cos(\theta(t)+\omega)+e\cos(\omega)\right),
\end{equation}
where $a_{\rm X} \sin i$ is the projected semi-major axis of the pulsar, $P_{\rm 
orb}$ is the 
orbital period, $e$ is the eccentricity, and $\omega$ is the longitude of 
periastron. The true anomaly $\theta(t)$ is found by solving Kepler's equation.

We check whether the spin period derivative obtained in this work, and 
also during previous analyses, could be explained as a consequence of the 
orbital motion of the pulsar along its binary orbit. Following 
Eq.\,\ref{eq:derivativeobs} we calculate the period derivative caused by the 
orbital Doppler effect assuming no intrinsic changes in the actual spin 
period and all the change we observe is only due to the orbital motion.
This derivative depends  on the orbital parameters of the binary system. 

Unfortunately,  the orbit of \sxp\ is not known. To roughly estimate a lower 
limit for the eccentricity we assume that the progenitor 
binary had a 
circular orbit  and zero kick velocity. Then, after the SN explosion:
\begin{equation}
\label{eq:SNexplosion}
  e = \frac{\Delta M}{M_1 + M_2 - \Delta M}
\end{equation}
\citep{Verbunt1995}, where $M_1$ and $M_2$ are the masses of the stars before 
the SN explosion 
and $\Delta M$ is the mass loss of the system during this explosion. 
\cite{Henault2012} estimated an evolutionary mass of $M_1\sim15$\,M$_\odot$ for 
the Be-companion of \sxp. Assuming that the NS in \sxp\ has the canonical mass 
of  
a NS  (we neglect the difference between baryonic and gravitational 
masses), i.e., $1.4$\,M$_\odot = M_2 - \Delta M$ \citep[e.g.,][]{Lattimer2014}, 
we estimate the mass loss to be $\Delta M > 14$\,M$_\odot$. 
 
Consequently, the eccentricity is expected to be close to unity. 
For our purpose we adopt a conservative approach and use as  
lower limit for the eccentricity $e>0.4$. Note, 
that lower eccentricity values for the orbit result in a higher 
probability for the spin period derivative to be intrinsic. We adopt $e<0.88$, 
which is the highest eccentricity known to date for a BeXB 
\citep[2S~1845$-$0242;][]{Finger1999}, as the upper limit for the eccentricity. 
Consequently, we assume the eccentricity to be uniformly distributed between 
0.4 and 0.88. Finally, we adopt an appropriate range of possible orbital 
parameters as follows:
\begin{enumerate}
\item $P_{\rm spin}=1062$\,s  as published by \cite{Henault2012} for simplicity.
\item $P_{\rm orb}=668\pm10$\,d (see Section\,\ref{subsec:orbitalper}).
\item $0.4\lesssim e \lesssim 0.88$
\item The longitude of periastron ($\omega$) is a geometric effect. 
Thus, we allow any value between -180$^\circ$ and +180$^\circ$ and 
assume a uniform distribution.
\item 
Following the third Kepler law, the semi-major axis of the orbit of the NS 
is $a_{\rm X}=1713.46$\,light-s. The orbital inclination is not known, 
therefore, we let it vary between $0$ and $90^\circ$.

\end{enumerate}

\begin{figure}
\begin{center}
\includegraphics[angle=0,width=1.0\linewidth]{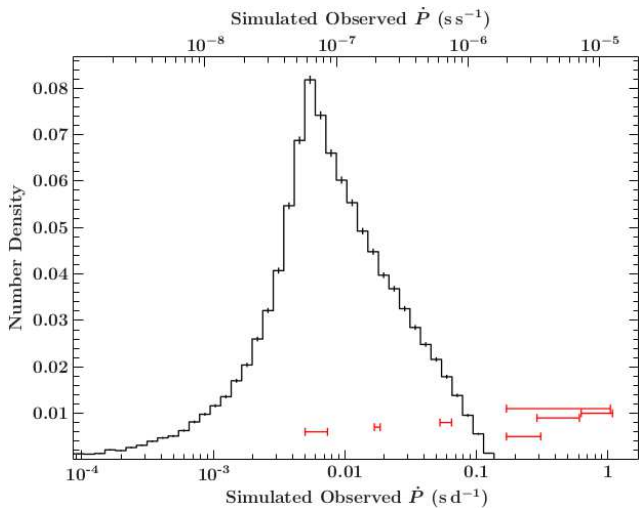}
\caption{Distribution of expected $\dot{P}$-values caused by orbital Doppler 
shifts 
according to the range of values for the orbital parameters estimated for \sxp\ 
(see Section\,\ref{sec:doppler}). The red horizontal lines indicate the position 
of the spin period derivatives given in Table\,\ref{tab:spin}. 
\label{fig:pdotorbit}}
\end{center}
\end{figure}

We performed 100000 Monte Carlo simulations where we calculated the spin 
period derivative choosing randomly the orbital parameters within the 
previously estimated ranges. The maximum of the simulated $\dot{P}_{\rm 
spin}(t)$ for each Monte Carlo run is binned into a histogram that results in a 
distribution of probability to observe a certain $\dot{P}_{\rm spin}$-value 
caused by the orbital motion of the binary. Fig.\,\ref{fig:pdotorbit} shows this 
distribution of probabilities. We assume pulse period derivatives out of this 
distribution to be intrinsic to the pulsar, while those which are within this 
distribution have different probabilities of being intrinsic depending on their 
position with respect to the distribution (see last column of 
Table\,\ref{tab:spin}).

\section{Discussion}
\label{sec:dis}

\subsection{The Be star decretion disc and the X-ray flux}
\label{sec:disc}

The H$\alpha$ line is the primary indicator for the presence of a decretion 
disc around a Be star and its $EW$ provides information about the size of 
this disc. In Section\,\ref{sec:obssalt} we presented \salt\ observations at 
different epochs which allow us to study the time evolution of H$\alpha$ and 
H$\beta$ emission lines and their $EW$. The non-split shape of Balmer lines and 
the time evolution of the absorbing column  
deduced from X-ray spectra indicates that we likely observe the Be disc face on. 

\citet{Reig1997} and, most recently \citet{Haberl2016} showed  a correlation 
between the orbital periods and the $EW$ of 
the H$\alpha$ line in BeXBs which follows a linear trend with larger H$\alpha$ 
emission for longer orbital periods. This correlation was explained in the 
framework of the truncation disc model.  The disc truncation is 
responsible for the so-called Type\,II X-ray outbursts, and inhibits the growth 
of the decretion disc \citep{Negueruela2001, Okazaki2001}. 
Interestingly, \sxp\ does not follow the H$\alpha$-orbital period correlation,
likely its  Be-disc is smaller than usual for its orbital period. 

As can be seen from Table\,\ref{tab:halpha} the equivalent width of H$\alpha$ 
in optical spectra of \sxp\ keeps  increasing over years, however it drops 
after the X-ray outburst, as can indeed be expected for a BeXB. This  change 
provides further confirmation that the NS interacts with the Be-disc during the 
outburst. 
However, the decrease in the disc size after the outburst is not dramatic, 
indicating that the disc of the donor star in \sxp\ is likely not truncated 
during the outburst.
%


\subsection{Evolution of the spectral parameters}

As shown in sections \ref{sec:xq} and \ref{sec:cho}, the basic X-ray spectrum 
of \sxp\ is an absorbed power law. This is similar to X-ray spectra of  other 
BeXBs in the SMC \citep{Haberl2008}. 

Our data set allows us to investigate how the X-ray spectrum of \sxp\ changes in 
the different luminosity regimes, i.e. during quiescence and post-outburst. As 
seen in Table\,\ref{tab:specfit} immediately  after the 
outburst the absorption column is strongly increased compared to the 
quiescence, and then decreases with time. At the same time, the 
spectrum initially softens (power-law index drops), and then hardens 
post-outburst. This points out to the 
accumulation of matter in the vicinity of the X-ray source towards and during 
the outburst, and a fast depletion afterwards.   

We tentatively interpret this behavior in the framework of a model that assumes 
that the outburst is triggered by the passage of the NS through the Be-disc. 
The orbit of the NS is probably inclined to the disc. 
According to this scenario, after the NS pierces the Be-disc, it is  observed 
through the disc material. As the time progresses, the column density  
decreases to its low quiescence value. Similar behavior is also observed  
in other systems, e.g.\ SXP\,5.05 \citep{Coe2015}. Future work, especially 
hydrodynamic modelling of winds from fast rotating Be stars, shall clarify 
whether the scenario we consider here is viable.

\subsection{Spin period evolution and magnetic field}
\label{sec:spinb}

\begin{figure}
\begin{center}
\includegraphics[width=0.9\linewidth]{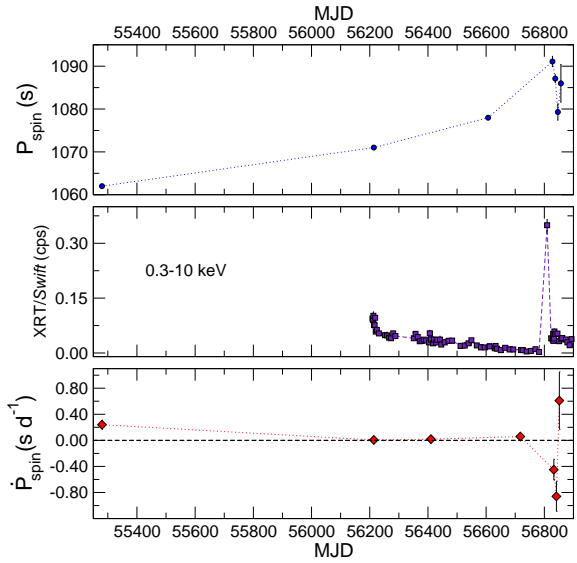}
\caption{{\it Top panel:} Temporal evolution of the pulse period of \sxp\  
(Table\,\ref{tab:spin}). 
{\it Middle panel:} XRT/\swift\ $0.3-10$\,keV X-ray light curve of \sxp. {\it 
Bottom panel:} \sxp\ spin period derivative as a function of time 
(Table\,\ref{tab:spin}). \label{fig:ppdotflux}}
\end{center}
\end{figure}

In this work we present new spin-period measurements for \sxp. With these new 
data we confirm that the pulsar appears to be spinning down continuously since 
it was discovered in March 2010 until at least, October 2013, which is the date 
of our quiescence \xmm\ observation. Our first \chandra\ 
post-outburst spin period (1091\,s; 2014-06-19) is longer than 
observed in October 2013 (1077.97\,s), consequently, the pulsar could even have 
been spinning down until some time before the outburst detected by \swift\ 
in June 2014 (see Fig.\,\ref{fig:ppdotflux}). However, we have observed for the 
first time a spin up for \sxp\ just after the outburst detected by 
\swift. We also observe how \sxp\ seems to recover its usual spin down in 
the last observation. To confirm this spin down we would need more observations.

Next we estimate the magnetic field of \sxp.
{\change The present day magnetic field of the NS in \sxp\ is not known. An
interesting proposal has been made by \citet{Fu2012} and \citet{Ikhsanov2012} 
and discussed later by other authors, that this compact object can have a
magnetar-scale field. In this case, the NS would belong to the hypothetical 
class of accreting magnetars proposed by \citet{Reig2012}. Besides \sxp, some 
other long period X-ray pulsars are suggested to be accreting magnetars  
\citep[e.g.][]{Sanjurjo2017}. Especially after the recent discovery 
of ultra-luminous X-ray sources with NSs the accreting magnetar 
idea gained further support \citep{Ek2015, Mush2015}. However, 
the possibility that Myr old accreting NSs have very strong magnetic 
fields questions the standard scenario of field evolution in NSs 
\citep{Pons2009,Turolla2015}

Our target,  \sxp, represents an extremely rare case of a relatively young NS 
with known age. This object provides unique insights to the NS magnetic field 
evolution. Therefore, we believe that despite the challenges related to the 
scarcity of the data it is of great importance to constrain its magnetic field.}

{\change There are various methods commonly applied for indirect estimates of 
the NS magnetic field based on its spin behavior \citep[e.g.][and references 
therein]{Shi2015,Angelo2017}. Such methods, based on older models of 
accretion, have been used in the past to derive strong magnetic fields for NSs 
in Be/X-ray binaries \citep[e.g.][and references therein]{Klus2014,Ho2014}.  
In the following we combine the spin period evolution and system parameters of 
\sxp\ to estimate the magnetic field strength of its NS. Our analysis is based 
on a recent model of quasi-spherical settling accretion developed by 
\citet{Shakura2012}.} 

\subsubsection{Spin period equilibrium hypothesis}

We start with a magnetic field estimate based on the hypothesis of spin 
equilibrium. {\change This assumption is often used to derive field estimates
in the case of X-ray pulsars.} 
It is supposed that in a  stationary situation spin-up and spin-down torques  
balance each other, and so the spin period is equal to the equilibrium value 
\citep{Davidson1973}. This quantity depends on several parameters of the system 
(accretion rate, properties of the accretion flow, parameters of the binary 
system, etc.). In particular, it depends on the magnetic field of the NS. 
Therefore, if all other parameters are more or less well known we can estimate 
the magnetic field using the following equation from \cite{Postnov2015} (see
details on the derivation of this formula in \citealt{ufn2013}):

\begin{equation}
P_{\rm eq}=940\,\mu_{30}^{12/11}(P_{\rm orb}/10\, {\rm d}) \dot M_{16}^{-4/11} v_{\rm w8}^4   \, {\rm s}.
\end{equation}
Here $v_{\rm w8}$ is the stellar wind velocity in units of $10^8$~cm~s$^{-1}$, 
$\dot{M}_{16}$ is the mass accretion rate 
in units of $10^{16}$\,g\,s$^{-1}$, 
and $\mu_{30}$ is the magnetic moment in units of $10^{30}$\,G\,cm$^{3}$ 
($\mu_{30}=1$ corresponds to $B=10^{12}$\,G on the equator for the 
standard $R_{\rm NS}\sim10$\,km; 
note, that for $R_{\rm NS}\sim15$\,km
\citep{bogdanov2013}, the field $B$ would be half an order of magnitude
smaller for the same $\mu$). 
This equation is applicable to systems 
with quasi-spherical accretion 
with $\dot M\lesssim10^{16}$\,g\,s$^{-1}$ 
\citep[so-called {\it settling accretion regime} 
which is applicable to X-ray pulsars 
with $L_{\rm X}\lesssim10^{36}$\,erg\,s$^{-1}$, see][]
{Shakura2012}. 

Substituting the parameters for \sxp\ we obtain:

\begin{equation}
\mu_{30}\approx 7 \left(\frac{\dot M}{5 \times 10^{15} \,
\mathrm{g\,s}^{-1}}\right)^{1/3} 
\left(\frac{v_{\rm w}}{200\,\mathrm{km\,s}^{-1}} \right)^{-11/3}.
\end{equation}

Assuming wind velocities $v_{\rm w}\sim200$\,km\,s$^{-1}$, 
which is typical for quasi-spherical winds observed from Be stars 
\citep{Waters1988}, the estimated magnetic field has 
a standard value of $B \sim 7\times10^{12}$\,G.
If we assume higher wind velocities the magnetic field would even be lower. 
Consequently, according to this estimate, 
we are not dealing with a present-day magnetar.

\subsubsection{Out of equilibrium hypothesis}

\begin{enumerate}
\item{Long-term spin down}
\parindent=0cm

We have seen that the pulsar might be out of equilibrium 
because we observe a continuous spin down over years 
(see Fig.\,\ref{fig:ppdotflux}). 
Therefore, we can use the data which correspond to the spin down of the pulsar 
to obtain another estimate of the magnetic field. 
Again, we use an equation from \cite{Postnov2015}:
\begin{equation}
\begin{split}
\dot P_{\rm sd}\approx 7\times 10^{-3} \left( \frac{\Pi_{\rm sd}}{7} \right)
\left( \frac{\dot M}{5 \times 10^{15} \, {\mathrm g\,s}^{-1}} \right)^{3/11} \\
\left(\frac{P_{\rm spin}}{1080~\mathrm{s}}\right) \mu_{30}^{13/11} \, \mathrm{s\,d}^{-1}
\end{split}
\end{equation}
Here $\Pi_\mathrm{sd}$ is a parameter of the theory of settling accretion, 
its typical value is $\sim$5$-$10. \citet[][and references therein]{Postnov2015} 
demonstrate that this parameter does not change much
from system to system. As we cannot derive it independently, we choose a
typical value  $\Pi_\mathrm{sd}=7$. The variation of this parameter might have a
very limited (less than a factor of two)  influence on the estimate of $\mu$.
From the observations of \sxp\ 
we can estimate an average $\dot P_\mathrm{sd}\approx 0.02$~s~d$^{-1}$.
Then we obtain $\mu_{30}\approx 3 $, which corresponds to $B\sim 3\times 
10^{12}$\,G. 
Taking into account many uncertainties in models and parameters, 
this result is in reasonable correspondence with the estimate obtained above 
from 
the equilibrium period. Note, that these values are in correspondence with the 
saturation value for magnetar field decay \citep{Pons2009, 
Gourgouliatos2014}.\\

\begin{table*}
\begin{center}
\caption{Main properties of the two established Be X-ray binaries 
sub-groups.\label{tab:bex}}
\begin{tabular}{lll}
\hline
& {\bf Transient} & {\bf Persistent}\\
\hline
{\bf Quiescence} $ L_{\rm X}$ & $\lesssim10^{33}$\,erg\,s$^{-1}$  & 
$\sim10^{34-35}$\,erg\,s$^{-1}$\\[0.2cm]
{\bf Outbursts} & Two different types & Relatively quiet systems\\
& Type I: Periodic with orbital period and  $L_{\rm 
X}\lesssim10^{37}$\,erg\,s$^{-1}$& Sporadic 
and non-predictable increases of $L_{\rm X}$\\
& Type II: Random with $L_{\rm X}\sim10^{38}$\,erg\,s$^{-1}$ & Increases of less 
than one order of magnitude\\[0.2cm]
{\bf Pulse period} $P_{\rm spin}$ & $\sim10$\,s & $\gtrsim200$\,s\\ [0.2cm]
{\bf Orbital period} $P_{\rm orb}$ & $20\lesssim P_{\rm orb}\lesssim100$\,d & 
$\gtrsim200$\,d\\ [0.2cm]
{\bf Eccentricity} $e$ & $\gtrsim0.3$ & $\lesssim0.2$ \\ 
\hline
\end{tabular}
\end{center}
\begin{flushleft}
{ NOTE:} The data presented in this table have been collected from the 
following papers:
\cite{Campana2002,Okazaki2001, Negueruela2001, Reig2011, Knigge2011, Cheng2014}
\end{flushleft}
\end{table*}

\item{Spin up during the outburst}
\parindent=0cm

Finally, we  discuss 
the spin up rate $\dot P_\mathrm{su}$ observed 
during the outburst detected by our \swift\ monitoring campaign. For this 
case we can neglect the braking torque. 
Then the equation for the evolution of the period reads:

\begin{equation}
\dot P_{\rm su}=P_{\rm spin}^2 P_{\rm orb}^{-1} \dot M
R_{\rm B}^2 I^{-1},
\end{equation}
where $I$ is the moment of inertia of the NS, we assume a standard
value $10^{45}$~g~cm$^2$ for it.
$R_\mathrm{B}$ is the Bondi radius. 
This equation provides an upper limit for the modulus
of $\dot P_{\rm su}$, but
realistic values are not much smaller \citep{lipunovbook1992}.

Note, that the maximum spin up happens 
during outbursts which appear close to the periastron. 
Then, we cannot take an average value for $R_{\rm B}$ which depends on $v_{\rm 
orb}$ 
and hence on the eccentricity. For larger eccentricities we obtain larger spin 
up. If we take $e=0.7$, $\dot M=10^{17}$~g~s$^{-1}$, stellar mass $15 \, 
M_\odot$ and $v_{\rm w}=200$\,km~s$^{-1}$, we obtain $\dot P_\mathrm{su}\approx 
0.5 $~s~d$^{-1}$. The observed average value for \sxp\ is $\dot P_{\rm su}\sim 
0.5$~s~d$^{-1}$. Then, we can conclude that the spin-up can be described by 
relatively highly eccentric orbits, which coincides with the estimation we have 
performed of the orbital parameters (see Section\,\ref{sec:doppler}).

\end{enumerate}

These simple {estimates} agree within uncertainties with those presented 
in Sec.\,3 and are in line with the theoretical expectations { for
magnetic field evolution of magnetars}.
Therefore, we can explain the spin period derivative observed 
in \sxp\ applying the theory of quasi-spherical accretion \citep{Shakura2012} 
without the need of an extremely strong magnetic field.
Future and in depth theoretical modelling of \sxp\ is necessary to fully 
verify this conjecture {\change  which is of general interest in the context of
yet poorly understood magnetic field evolution of NSs}.

\subsection{On the classification of \sxp}

BeXBs are classified into two broad subgroups: transient BeXBs and persistent 
BeXBs \citep[][]{Reig2011}, which may overlap in part of their properties. 
In Table\,\ref{tab:bex} we summarize the key characteristics of these sub-groups.

\sxp\ has properties that are shared by both of these sub-groups. Its 
large X-ray luminosity in quiescence (see Table\,\ref{tab:lum}),  the 
long binary orbital period  $P_{\rm orb}=668\pm10$\,days, and its long pulse 
period  $P_{\rm spin}\sim1070$\,s, are typical for persistent BeXBs. 
Also, to date no known BeXB system containing a NS with $P_{\rm 
spin}>100$\,s has an eccentricity above 0.5 \citep{Townsend2011}.
On the other hand, the Type I outburst reported in this work, although moderate, 
is a characteristic of transient BeXBs. 
Therefore, the properties of \sxp\ suggest that systems exist which can be classified
as member of both groups and that transient and persistent BeXBs are not completely 
separated sub-groups.

\section{Summary and Conclusions}
\label{sec:conclusion}

In this work we present a coordinated optical and X-ray study of the
X-ray pulsar \sxp\ during quiescence and outburst.  The main results of our study 
are:

\begin{itemize}
\item[$-$] We observed a third Type\,I outburst in \sxp. The outburst 
occurred at a regular interval from two previously observed.  The binary 
period is constrained to $P_{\rm orb}\approx 668\pm10$\,d. Type\,I outbursts 
are characteristic for transient BeXB. Thus \sxp\ shares observational 
characteristics with persistent as well as transient BeXBs.

\item[$-$] The profiles of the H$\alpha$ line indicate that we likely observe 
the Be-disc face on, while the temporal evolution of its $EW$ shows that the 
disc is not truncated in the aftermath of the outburst. 

\item[$-$] We confirm the earlier reports on the presence of a black body spectral 
component in the quiescent X-ray  spectrum of \sxp.

\item[$-$] Our analysis of the pre- and post-outburst spectra reveals an abrupt 
increase in the absorption column just after the outburst. Taking into 
account the H$\alpha$ line profile, we propose that the disc of 
the Be star is observed face on, while the orbit of the NS is inclined 
to the disc. 

\item[$-$]  Studies of the NS spin evolution reveal that  the NS had been 
intrinsically spinning down at least from March 2010 until 
October 2013. We report a brief post-outburst spin up of the X-ray pulsar.

\item[$-$] We estimate the magnetic field for \sxp, assuming different 
hypotheses for the spin period evolution (e.g., close to equilibrium, out of 
equilibrium) and the quasi-spherical accretion model \citep{Shakura2012}. From 
these calculations we find no need for an extremely high magnetic 
field at present day for the pulsar in \sxp.
\end{itemize}

\section*{Acknowledgments}
The authors thank the referee for useful comments and suggestions that helped 
to improve the paper. 
SBP thanks Profs. N.I. Shakura and K.A. Postnov for discussions.
AGG and LO are supported by the Deutsches Zentrum f\"{u}r Luft und 
Raumfahrt (DLR) grants FKZ 50 OR 1404 and FKZ 50 OR 1508. 
LO acknowledging  partial support by the Russian Government Program of 
Competitive Growth of Kazan Federal University.
MK acknowledges funding by the Bundesministerium f\"ur Wirtschaft und 
Technologie under Deutsches Zentrum f\"ur Luft- und Raumfahrt grants 50 OR 1113 
and 50 OR 1207. SBP is supported by the Russian Science Foundation grant 
14-12-00146.MPES received funding through the Claude Leon Foundation 
Postdoctoral Fellowship program and the National Research Foundation.


\newpage

\begin{appendix}
\section{LOG of SALT observations}

\begin{table}
\begin{center}
\caption{Log of optical RSS/\salt\, observations obtained with the VPH PG2300 
grating.\label{tab:opticalobservations}}
\begin{tabular}{ccccc}
\hline
\#  &  MJD & Date & $\Delta\lambda$ (\AA) \\
\hline
1 &  56827.80 & 2014-06-20 & $6082-6925$ \\
2 &  56835.49 & 2014-06-27 & $6082-6925$ \\
2 & & 2014-06-27 & $3835-4924$\\
2 & & 2014-06-27 & $3835-4924$\\
3 &  56847.30 & 2014-07-09 & $6082-6925$ \\
3 & & 2014-07-09 & $3835-4924$\\
3 & & 2014-07-09 & $3835-4924$\\
4 & 57694 & 2016-11-02 & $6082-6925$ \\
\hline
\end{tabular}
\end{center}
\begin{flushleft}
{ NOTE:} The wavelength coverage of the grating changes with the input 
angle. 
The spectra covering the H$\alpha$ region ($6082-6925$\AA) were obtained with 
the angle of $48.875^\circ$, while the spectra covering the H$\beta$ and 
H$\gamma$ region ($3835-4924$\AA) were obtained with the angle of $30.5^\circ$.
\end{flushleft}
\end{table}

\end{appendix}

\end{document}